\documentclass{article}
\title{Experiment and the Pursuit of Ugly Models}
\author{Martin King\footnote{Munich Center for Mathematical Philosophy, Ludwig Maximilian University of Munich}
}
\usepackage{natbib}
\usepackage{graphicx}
\usepackage{enumitem}
\usepackage{amsmath,amssymb}
\graphicspath{ {../../images/} }
\usepackage{lineno}
\begin{document}

\maketitle

\begin{abstract}
Scientists do not merely choose to accept fully formed theories, they also have to decide which models to work on before they are fully developed and tested. Since decisive empirical evidence in favour of a model will not yet have been gathered, other criteria must play determining roles. I examine the case of modern high-energy physics where the experimental context that once favoured the pursuit of beautiful, simple, and general theories now favours the pursuit of models that are ad hoc, narrow in scope, and complex; in short, ugly models. The lack of new discoveries since the Higgs boson, together with the lack of a new higher energy collider, has left searches for new physics conceptually and empirically wide open. Physicists must make use of the experiment at hand while also creatively exploring alternatives that have not yet been explored. This encourages the pursuit of models that have at least one of two key features: i) they take radically novel approaches, or ii) are easily testable. I present four cases, neutralino dark matter, the relaxion, repulsive gravity, and models of lepton flavour universality violation, and show that even if they do not exhibit traditional epistemic virtues, they are nonetheless pursuitworthy. I argue that experimental context strongly determines pursuitworthiness and I lay out the conditions under which experiment encourages the pursuit of ugly models.
\end{abstract}

\textbf{Keywords:} pursuitworthiness; philosophy of experiment; theoretical virtues; values in science; philosophy of physics; history and philosophy of physics; philosophy of particle physics; supersymmetry; relaxion; repulsive gravity

\section{Introduction}

Philosophers of science have long tried to identify the characteristics of good scientific theories---the theories that one should rationally opt for among rivals. 
This includes epistemic criteria, such as internal consistency and the scope of the theory's fit with empirical data. 
But philosophers have also recognized that some important criteria are perhaps not strictly epistemic, but cognitively beneficial for those working with the theories, such as the theory's simplicity. 
Of course, scientists do not simply choose to accept theories after they are fully formed, they also have to decide which to work on before they are theoretically developed and empirically tested. 
A growing literature on the pursuitworthiness of theories has emerged stemming from early work by \citet{Laudan1977-LAUPAI} who distinguished between the context of acceptance and the context of pursuit.
These decisions of pursuit are inevitably made on the basis of considerations that are more pragmatic and heuristic and, I will argue, are strongly influenced by the experimental context.

In order to show this, I will examine the case of contemporary high-energy physics (HEP). 
The changing experimental context of HEP in recent years makes clear the influence on criteria of pursuit. 
For decades, the Standard Model (SM) has guided much of the experimental and theoretical research in HEP and since the Higgs boson was discovered in 2012, the SM is in some sense complete. 
Physicists have long been devising alternative theories and higher-energy completions of the SM, but no significant evidence in favour of any of these models has been uncovered at the LHC. 
Models that predicted new particles that would be clearly visible in the data as sharp resonances above
 the SM background simply did not manifest. 
When it comes to new physics models, the low-hanging fruit is gone. 
``Growing emphasis is given to alternative scenarios and more unconventional experimental signatures where new physics could hide, having escaped traditional searches''\citep[p.~113]{Ellis:2691414}. 
The models that remain in the available parameter space and are the subject of most research can be complex, difficult to calculate, ad hoc, have small empirical scope, are not unifying, not very plausible, etc. 
In short, the models on the table are \textit{ugly}---they do not exhibit traditional theoretical virtues.
What I argue is that these ugly models are nonetheless pursuitworthy since they exhibit a set of criteria that are important in our current experimental context.

I use the term `ugly' in order to stress the deviation from the conventional standards of scientific beauty, such as elegance, accuracy, simplicity, etc. 
I am highlighting the pursuitworthiness of these models, despite their ugliness. 
In much the same way, one can appreciate some pieces of modern art that do not exhibit conventional beauty.\footnote{A different lesson one could take from these case studies is that ad hoc models are beautiful too (in a non-conventional sense).} 
The argument of the paper does not properly concern aesthetics in science, or address the question of whether there are epistemic reasons to pursue aesthetically pleasing theories (for that one can refer to \citep{Ivanova2020Aesthetics}). 
I am rather noting that a conventional notion of beauty exists and that the pursuit of these models comes apart from that.
It should also be noted that there is not a particular threshold under which we have crossed that the models are now ugly, but rather that they are \textit{uglier}, as I will stress in Sec.~\ref{sec:newphys}.\footnote{One could make the case that the SM, against which these models are being compared, is itself quite ugly. After all, it has an ad hoc mechanism of spontaneous symmetry breaking, unexplained neutrino masses, three fermion generations, accidental symmetries, many free parameters, etc.}

I will highlight two important experimental conditions in the state of HEP: that there has been no evidence in favour of new physics with our current experiments; and second that there will be no new higher-energy experiment for some time (more on this in Sec.~\ref{sec:newphys}). 
These conditions lead to two important guides for pursuit in HEP: i) radical novelty; and ii) immediate testability.
The increasing importance of novelty stems from a sense that physicists have already exhausted the avenues that were most appealing in terms of epistemic virtues---radically different approaches are needed. 
More than mere predictive novelty is meant.
What is particularly valuable is a new kind of theoretical idea that allows physicists to approach problems in a new way, with new tools and insights.

Immediate testability enters because with no new higher-energy collider on the horizon, there is no possibility of simply turning on a machine and seeing evidence of a new physics theory Beyond the Standard Model (BSM). 
There is limited empirical progress to be expected in working on theories outside of achievable energies.
In order to maximize chances of discovering new physics, physicists have had to get very clever: they have had to re-evaluate long-cherished guiding principles and the very foundations of quantum field theory (QFT); and they have had to get very creative with model-building, experimentation, and data analysis.
This points not to a decline in standards, but rather indicates that a different set of criteria are playing a stronger role in determining pursuit.

The paper will begin by discussing epistemic virtues and the shift to pursuitworthiness in Sec.~\ref{sec:epistemic}.
Then in Sec.~\ref{sec:newphys}, I look at four ugly models that are pursuitworthy, but do not exhibit the traditional virtues of theories. 
Finally, I distil two desiderata (novelty and testability) from the actual cases whose importance is not evident on extant accounts of pursuit in Sec.~\ref{sec:pursuit}. 
Detailed descriptive accounts, such as \citep{Franklin1993-FRADPA-3} and decision-theoretic models \citep{Nyrup2015-NYRHER} do not stress or capture the experimental context-dependence that I wish to highlight.
As such, the aim of the paper is not to present a fully fleshed-out account of pursuitworthiness. 
In fact, a general account may not be possible given the context-dependence I argue for, in agreement with arguments by \citet{Shaw2022-SHAOTV-2}. 
Ultimately, what I hope to show is that the context in which research takes place strongly determines considerations of scientific pursuit---it is the way particle physics is now that leads us to novelty and testability as important criteria.

\section{Epistemic Virtues and Pursuitworthiness}		\label{sec:epistemic}

The debate is often traced back to \citet{kuhn74}, who outlined the most important criteria for theory choice---accuracy, consistency, simplicity, scope, and fruitfulness\footnote{Kuhn's list and arguments sparked a long history of debate about the status these criteria and other criteria, opening what can be described as a new demarcation problem \citep[see for example]{douglas2009,lacey99,laudan84,longino95,mcmullin85}, but that is not the avenue this paper will pursue.}.
He included these not because they are exhaustive, but "they are individually important and collectively sufficient to indicate what is at stake" (p.~357).
He sought to reassure people of the objectivity that was possible within the framework developed in \textit{Structure} \citep{kuhn62}, but noted that how these criteria are assessed is still a subjective matter. 
Two scientists may agree that simplicity is important, but they may take simplicity to mean different things and they may disagree on how much it ought to weigh against other criteria like accuracy or scope. 

An important contribution to the debate that I wish to pick up on was made by Douglas. 
\citet{douglas2009,douglas2013} proposes a finer-grained account of values in science in part to show that they are not necessarily in tension with each other. 
She does this by making two important distinctions: that between minimal criteria and desiderata, and that between values as applied to theories themselves and as applied to theories in relation to evidence. 
There are two such minimal criteria for a scientific theory: internal consistency and empirical adequacy; internal consistency is a property of the theory in itself and empirical adequacy is a property of the theory in relation to evidence. 
She separates other cognitive values into pragmatic (which apply to the theory in itself) and epistemic (which apply to the theory in relation to evidence).
The latter two are mere desiderata on her account and include such values as simplicity and breadth of scope.
Thus, it is pragmatically beneficial if a theory is simple in itself, but it is epistemically valuable if the theory is simple in relation to evidence. 
What will be important for our discussion is that even so-called minimal epistemic criteria are not necessary for pursuit.

The difference between acceptance and pursuit is not simply a matter of a lower threshold for the same criteria. 
If it were, models that have not yet been fully developed would would always lose out in comparison with established theories. 
One would then always be pursuing the best developed and empirically established theories, which misses the point of pursuit and misses an important scientific practice.
As Laudan puts it: ``if one insists (as virtually all philosophers of science do) that standards for accepting a theory should be pretty demanding epistemically, then how can it ever be rational for scientists to utilize new theories which (in the nature of the case) will be likely to be less-well tested and well-articulated than their older and better-established rivals?'' \citep[p.~232]{Laudan1996-LAUBPA-3}.
In order to rationally opt to work on some models and hypotheses before there is conclusive evidence and before all the conceptual problems have been straightened out, one must have a different set of criteria on which this choice can be made.\footnote{Some recent work on pursuit in physics has focused on the comparison of dark matter and modified gravity research programs \citep{martensking,Wolfduerr}.} 

Pursuitworthiness, as an explicit concept, goes back to \citep{Laudan1977-LAUPAI} and \citep{McMullin1976-MCMTFO-3} and significantly developed by \citep{Achinstein1993-PETHTD} and \citep{Franklin1993-FRADPA-3}. 
The idea is to shift the focus from the justification of acceptance to the justification of pursuit, where decisions are not necessarily about truth or realism, but can focus on short timescale decisions. 
For \citep{Laudan1977-LAUPAI}, pursuit can include refining a hypothesis\footnote{Franklin often speaks of pursuing hypotheses, others speak of theories or of models. I take the line between model and theory to be more-or-less continuous, but one could reserve the term theory for a well-developed and broadly empirically confirmed model. Thus, theories are the kinds of things that would be subject of theory-choice decisions and models the subject of decisions of pursuit. One could of course pursue just a hypothesis, which may not yet have much mathematical structure or conceptual content and may only be about the value of some observable.} empirically, developing it theoretically by solving conceptual problems, and so on. 
And importantly, one can be justified in working on theories with serious conceptual or empirical problems, if, in line with his account of scientific progress, the theories have resources to potentially solve otherwise unsolved problems. 
This idea is echoed by \citet{Franklin2016-FRATRA} who says that ``a hypothesis can act as a stimulus for further work even if one were skeptical of both the hypothesis and the evidence supporting it'' (p.~82) and that ``belief in the truth of a hypothesis or in the experimental results is also not a requirement of further theoretical work,''\citep[p.~106]{Franklin1993-FRADPA-3}.
It is not the case that one should pursue the theory that is best supported by evidence, has the largest scope, or even restrict oneself to theories that are consistent. 
These are the precisely the points at which considerations of pursuit pull apart from those of acceptance. 
The ``basic criteria that any scientific work must meet'' \citep[p.~92]{douglas2009} are not yet met in the context of pursuit.
Decisions to pursue are less constrained by these basic epistemic considerations, which allows many more features of the theory and the context to play leading roles in the decisions.
This also brings with it a worrying openness in descriptions of pursuitworthiness assessments. 
Let us look at a few accounts of how decisions of pursuitworthiness are made. 

\citep{mckaughan2008} argues that Pierce's account of abduction and Inference to the Best Explanation (IBE) was really about pursuitworthiness. 
He articulates Pierce's view as follows: ``if we estimate that testing the hypothesis will be easy, of potential interest, and informative, then we should give it a high priority'' \citep[p.~457]{mckaughan2008}. 
This is a very reasonable view, but one could easily imagine that these conditions are not all satisfied, or are satisfied by different models to different degrees. 
This alone is of little help to those hoping for a normative account of pursuit. 
As another example, \citet{Franklin2016-FRATRA} takes a detailed historical look at how decisions were made to pursue empirically and theoretically dubious models (in this case that of a fifth fundamental force). 
He sums up his analysis as follows: 
\begin{quote} 
	I suggested several reasons why a scientist might choose to further investigate, or to pursue, a hypothesis. These included the interest and importance of the hypothesis; its plausibility, based on existing evidence, on its resemblance to other successful theories, or on its mathematical properties; the fact that it fit in with an ongoing research program; and its ease of test, in which I include the conceptual simplicity of the test, which differs from the technical experimental details of the test, which might be quite complex; and whether or not the experiment can be performed with either existing apparatus or with small modifications of it, or with a relatively modest investment in a new apparatus.
\end{quote}
\begin{flushright}
	{\citep[p.~80]{Franklin2016-FRATRA}}
\end{flushright}
This is a rather long and permissive disjunction that can hardly be wrong. 
There are surely many reasons that a scientist could consider when deciding to pursue a given hypothesis and I believe that Franklin is quite correct in giving much place to the experimental side of the decision. 
Although, it is almost certainly true, the above quote is descriptive and less helpful in application than one might hope. 
Other accounts have provided more detailed breakdowns of how decisions of pursuit are, or ought to be, made.

One such is in given in \citep{Nyrup2015-NYRHER}, where the problem to be solved is put into a decision-theoretic framework. 
He says ``Scientists need to justify which hypotheses are worth investigating in order to prioritize their resources. Justifying pursuit is, essentially, a decision-theoretic problem of how to optimize the epistemic output of science'' \citep[p.~753]{Nyrup2015-NYRHER}.
The idea is to model the decision to pursue a hypothesis as a weighted sum of factors based on the expected epistemic value (EEV) of pursuing that hypothesis. 
One can then use the formalism of probability calculus to obtain rigorous results. 
Decision theoretic models in general can be informative in that separating out relevant factors and examining the role different weights would play can tell us a lot about how kinds of decisions can be affecting by the information that goes into them. 
For any given decision, calculating with real values of course will not be possible, so it is of limited applicability.
The messy details about how one comes to the number that feed into the calculation are precisely what most are interested in when discussing pursuitworthiness. 
For example, I am most interested in the influence of the ease of test and the novelty of the hypothesis, but there is no place for any such values, they must simply be folded in to the estimated value of pursuing the hypothesis.
My point in bringing up a formal account of pursuit is to express some concerns about the success of any such endeavour. 
This paper rather highlights the changing weightings of criteria of pursuit and their experimental context dependence. 
There is still a normative picture here, a rational story to tell about what is pursued, but it is open-ended and pluralistic. 

The general idea of weighing costs and benefits of choosing to pursue certain research lines rather than others is surely an accurate description of the choices scientists must make. 
A recent paper has proposed a `meta-methodological' framework for this that considers the weighing of cognitive benefits and costs without a calculus and this seems a more fruitful avenue than a toy formal model \citep{duerr2025rationallywarrantedpromisevirtueeconomic}. 

\section{Hunting for New Physics}		\label{sec:newphys}

The 27km long LHC at CERN began operation in earnest in 2010, colliding protons at an unprecedented 7~TeV. 
On July 1, 2012, two of the large experiments at the LHC, ATLAS and CMS, announced the discovery of a Higgs-like boson at 125~GeV. 
Early on, there was still room for this particle to exhibit non-SM-like properties, but as more analyses were conducted, it looked more and more like the `vanilla' Higgs of the SM, and the door began closing on new physics. 
The LHC, now operating at 14~TeV during Run 3, will soon be undergoing a major upgrade to higher luminosity (rate of collisions), which should be competed around 2028.
While this will allow an order of magnitude more data than the nominal LHC runs, expectations of discovering new particles are low. 
The stated physics goals of the upgrade do include two particular avenues for discovering new physics: stops, and neutralinos and charginos (see \citep{Schmidt_2016} for more details). 
While the additional data will greatly increase sensitivity and help increase the mass range that can be probed, the space is already heavily constrained. 
ATLAS has managed a 95\% exclusion limit on chargino-neutralino pairs up to around 850~GeV and stop pair production is excluded up to around 1~TeV \citep{aad2023,Williams:2891456} and the High-Luminosity LHC (HL-LHC) will push these beyond 1.5~TeV \citep{RUHR2016625}.

There are plans for the 90km long Future Circular Collider (FCC) at CERN that could achieve energies up to 100~TeV, but it is still in the proposal phase.
A feasibility study will be completed in 2025 and, if all goes well, the FCC high-energy phase could be operational in the mid-2070s (though a $e^+e^-$collider in the FCC could be operational in the late 2040s). 
The FCC's conceptual design report states: ``With the Higgs boson discovery, the standard model seems complete and its predictions have no more flexibility beyond the uncertainties in the theoretical calculations and in the input parameters\ldots The agreement between the predicted and observed W, top and Higgs masses, and the null result of experiments at colliders so far, are an indication that either the new physics scale is too high and/or the pertaining couplings are too small\ldots As a result, the next accelerator project must allow the broadest possible field of research'' \citep[p.~292]{Abada2019FCCee}.
There are two large sources of trepidation: whether or not this is the best use of the 15 billion Francs this would cost; and whether or not this is the best target of the physics community's efforts.
These too are important matters of pursuit, but far outside the scope of this paper, whose target is the features of particular models.
Arguments that helped make the case for the LHC, in particular naturalness arguments for the need to discover the mechanism of electroweak symmetry breaking near the electroweak scale and for new physics at low energies (in particular supersymmetry), are now regarded with suspicion. 
They are precisely the guides and assumptions being reevaluated in face of the lack of new discoveries.
There are no widely accepted arguments that new physics should be discovered below the achievable 100~TeV---the SM could be a perturbatively accurate description of nature up to the Planck scale ($10^{18}$~GeV). 

Part of an approach to find new physics elsewhere has involved shifting focus from the high-energy to the low-energy frontier \citep{jaeckel2010}, emphasising the prospects of discoveries with small experimental searches for axions (and axion-like particles). 
Another emphasis has been on finding long-lived particles the large detectors at the LHC are not sensitive to by building a small detector further down the tunnel.
This experiment (FASER) has provide constraints on dark photons \citep{CERN-FASER-CONF-2023-001}, and observed for the first time neutrinos produced at collider experiments \citep{Abreu_2023}. 
There are however, no hints of new physics here either. 

The lack of new physics discoveries has changed the field of particle physics in a number of significant ways. 
The European Strategy Update stresses that times have changed since the Higgs discovery:``In ten years of physics at the LHC, the particle physics landscape has greatly evolved''\citep[p.~1]{Benedikt:2653673}. 
One is that many physicists are increasingly turning to model-independent methods, to signature-based studies, and to models that would not be pursued in a different experimental context. 
To put it in a rather pessimistic way: physicists are scraping the bottom of the barrel for models, trying to get the most they can out of the experiments they have on hand. 
However, one never knows where a breakthrough will come from or when a small discrepancy will ultimately lead to a crucial test of a new theory. 
Hunting for new physics requires a multi-pronged approach involving among other things precision measurements, model testing, and pure theory. 
Physicists are working on uglier models in general and they should be, given the current open-ended context, but of course not all of them. 
One may describe the situation by saying that there are shifts in the division of cognitive labour. 
Where once there may have been an overwhelming number of theorists working on string theory and supersymmetry, there are now fewer, and there are more working on the remaining LHC-accessible models, even though they are uglier. 
This is a reasonable response to the situation. 

While I am stressing the context dependence of pursuit, it should also be noted that the context can rapidly change, for example if there were an unexpected discovery.
It may also change slowly with the passage of time as we approach the completion of a new experiment (perhaps the FCC), which may trigger particular theory work, modelling building, and phenomenology. 
The ugly models I am calling pursuitworthy here may no longer be as pursuitworthy as other perhaps more beautiful models.

\subsection{Ugly Models}		\label{sub:uglymodels}

In this section, I present four ugly models of BSM physics that are nonetheless pursuitworthy: neutralino dark matter; the relaxion; the aptly-named repulsive gravity; and models of lepton flavour universality violation. 
I choose four to highlight different problems, namely ad hocness, consistency, implausibility, and narrowness of scope, while drawing attention to the two features that make them most pursuitworthy regardless, novelty and testability.

\subsubsection*{Neutralino Dark Matter}		\label{subsub:nmssn}

The first model is the singlino-like neutralino model of dark matter, which is a next-to-minimal supersymmetric model \citep{Ellwanger_2010}. 
The model represents a class of solutions to a problem of the SM, in this case the dark matter problem.\footnote{The basic problem is that it seems that visible matter along with known physics falls drastically short of being able to reproduce basic observable elements of the cosmos. One can find more about this from a philosophical point of view in \citep{MillsBoyd2023-MILPOA-7}.}
The kind of solution it represents is weakly interacting massive particle (or WIMP) dark matter, where some new matter field is introduced, often in a new larger symmetry, which is light and stable, so that other particles can decay into it and then it will stick around and float about the universe. 
A particle is a DM candidate if it is neutral and interacts at most only weakly with other matter, so that we would not have seen it in our precision experiments to date, but has mass and so will have the gravitational influence on matter that we seem to see in astronomical observations. 
If one wants to study a model of dark matter, WIMP solutions are a perfectly reasonable avenue to pursue. 

Perhaps the best way to understand the some of the relevant parts of the model is to see how the model was arrived at.
The basic WIMP idea is to add a new particle to account for missing mass in the universe and we might think that SUSY is attractive for a variety of reasons. 
We might try to search for minimal supersymmetry (MSSM), which has four neutralinos, the lightest of which ($\widetilde{N}^0_1$) is a DM candidate. 
The model is theoretically capable of supporting predictions and is even LHC-accessible, so it can be empirically investigated. Further, as a supersymmetric model it could possibly solve many problems of the SM in one fell swoop, potentially unifying the coupling constants of the fundamental forces at high energies and providing a solution to the Higgs naturalness problem as well.  
But things starts to get ugly when we modify and constrain it on the basis of it not having shown up in our experiments.

There has been no evidence of the MSSM at the LHC. 
The energy scale at which SUSY may dwell can be pushed back by compromising features of the model. 
For example, a major benefit of the model was its ability to solve the Higgs naturalness problem, but as the SUSY scale get heavier, the hierarchy problem that was alleviated with loop corrections, begins to grow again and we are left with a `little hierarchy problem' between the electroweak and DM scales.
Important for our DM model in particular, the MSSM parameter space combined with DM relic abundance and direct detection data from LUX and XENON1T, is effectively excluded \citep{Abdallah_2016,Akerib_2017,Aprile_2018,Guchait:2020wqn}. 
If a preferred model's parameter space is getting too small, the obvious next step is to expand the parameter space, i.e. move to the less minimal model, viz. next-to-minimal SUSY (NMSSM). 
This model adds a Higgs singlet to the MSSM's extra Higgs doublet. 
This is an important addition as it increases the singlino and neutralino states of the model, changing its phenomenology and requiring new searches.
When this fails to turn up any evidence in searches, one can simply move to the next-to-next-to-minimal SUSY model and try again. 

This model is very ad hoc.
The model is less simple, more calculationally involved, requires an even larger number of hypothetical particles, each specified by many new free parameters. 
Some of these particles, including seven non-SM-like Higgses, are highly constrained by existing data \citep{CMS:2024yiy}.  
While many might give SUSY models in general high plausibility, this model resides a tiny, but not excluded, area of the parameter space. 
If one wants to examine the viability of such a model, one can numerically scan the parameter space to see if it is worth investigating. 
Elements of the parameter space are sampled and combinations are calculated to see if they remain in the empirically available parameter space. 
For NMSSM, there were only 2500 possible combinations values out of 25 million sampled in 2018 \citep{mou2018light}. 
One only looks at this model because other models have been ruled out and one only looks at these specific exotic regions of the parameter space because every where else has been excluded.

It is a remote possibility, but that is enough to make it pursuitworthy in the context where not much else is available.
Searching for DM candidates is still ``one of the most exciting and challenging programs'' of HEP \citep[p.~2]{Guchait:2020wqn} and is treated by CMS and ATLAS as a `high-priority' analysis \citep*{Abercrombie_2020}. 
The fact that it can be probed with existing apparatus and the parameter space constrained means that there is something to be learned in performing these analyses. 
There is also a rather significant payout if any part of the model were to be suggested by the data.
Getting a hint of supersymmetry or of a dark matter candidate would be a massive payout for a relatively low cost of continuing to probe accessible parameter space.

\subsubsection*{Relaxion}		\label{subsub:relax}

The second model represents a class of solutions to the Higgs naturalness or hierarchy problem that involves no new dynamics at the weak scale.\footnote{The Higgs naturalness problem is that the Higgs is a scalar field and in Quantum Field Theory one would expect to find the field around the energy level of the cut-off of the theory. But for the SM, this is (as far as we know) at the Planck scale. If the Higgs is light (and in this context 125GeV is light), then an extremely delicate cancellation of quantum contributions to thirty orders of magnitude is required. There is quite some debate about the best way of posing the naturalness problem and I will not go into this here. For an overview on the relevant positions and a philosophical analysis, see \citep{Harlander2019-HARHNA-2}.}
The relaxion is an alternative to the kind of solutions that SUSY provides and which may be more promising as the LHC precision data rules out models up to and beyond the TeV scale where new physics limit the amount of fine tuning the SM requires. 
The idea is to dynamically relax the electroweak scale via cosmological evolutions. 
In this model, an axion (more specifically, an axion-like particle) is added that couples to an inflaton (a field of cosmological inflation).
This can create a slow rolling periodic potential that `naturally' selects a light Higgs mass when the Higgs vacuum expectation value (vev) becomes non-zero as the relaxion stabilises after it passes a critical point.

\begin{figure}	\label{fig:relaxion}
	\includegraphics[width=\textwidth]{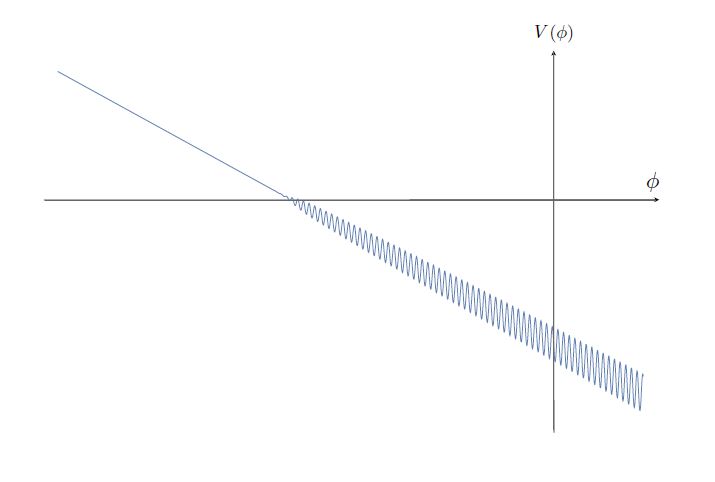}
	\caption{A 1-dimensional depiction of the relaxion's rolling periodic potential $V(\phi)$.}
\end{figure}

The characteristic rolling potential of the relaxion is depicted in Fig.~\ref{fig:relaxion} taken from the original paper proposing the solution in 2015 \citep{Graham2015}. 
The Higgs vev is selected when the slope of the linear scanning term balances with that of the periodic potential, so-called dynamical stopping. 
The vacua, the troughs in the potential, become classically stable near the weak scale with lifetimes much longer than that of the universe. 
The authors describe the model as follows: 
 \begin{quote}
 ``During inflation, $\phi$ will slow-roll, thereby scanning the physical Higgs mass. At some point in the $\phi$ potential, the quadratic term for the Higgs crosses zero and the Higgs develops a vacuum expectation value. As the Higgs vev grows, the effective heights of the bumps, $\Lambda^4$, in the periodic potential grow. When the bumps are large enough they become barriers which stop the rolling of $\phi$ shortly after $m^2_h$ crosses zero. This sets the Higgs mass to be naturally much smaller than the cutoff.''
 \end{quote}
 \begin{flushright}
 	\citep[p.~3]{Graham2015}
\end{flushright}
This is a radically novel approach to the naturalness problem that intrigued and inspired many physicists to work on variants of the model, as Franklin noted in some cases of pursuit, even though there were conceptual and evidential issues with the model. 

Just like the Higgs itself, the model is very ad hoc: the model is introduced to solve a particular problem without any empirical support. 
But beyond its ad hocness, the model as it was first proposed has a number of empirical and conceptual problems. 
While the model attempts to solve the naturalness problem, it is not even \textit{technically natural} (in the sense of \citep{tHooft1980}), since it breaks a gauge symmetry. 
The axion was first introduced as a potential solution to the strong-CP problem \citep{PhysRevLett.38.1440,PhysRevLett.40.279}, but here the particle is not a true axion (only axion-like) and in many models actually contributes to the strong-CP problem.
In short, a Higgs fine-tuning has been replace with a CP-fine-tuning problem, but at least it is orders of magnitude smaller. 
The aspect I would like to highlight is that the model is also inconsistent with cosmological observations and many accepted cosmological models. 
The mechanism for stopping the relaxion requires a long period of inflationary expansion so that the whole range can be scanned.
Rather than the 50 or 60-odd e-foldings\footnote{The amount of time required for growth by a factor of $e$.} one would expect on most cosmological models, in the minimal relaxion model, one needs $10^{33}$ or so\citep{Germ_n_2023,Nelson_2017}.

More work has been done on the model in order to alleviate some of these problems.
The relaxion's CP problem can be averted in the so-called `landscape relaxion' \citep{Nelson_2017}, in which the relaxion populates a landscape of universes, there is a noted temperature-dependence of the axion mass, which implies that the QCD contribution to the potential becomes much stronger after the inflationary period when temperatures drop well below the QCD scale. 
This model is then even more complex and ontologically richer with a landscape of universes, and still involves a finely-tuned cosmological constant. 
The gigantic e-folding number can be reduced, for example, by replacing the original dynamical stopping mechanism with a thermal stopping from particle production \citep{Hook_2016}, or from Hubble friction \citep{You_2017}, or from double-scanning with additional scalars \citep{Espinosa_2015}.
A stochastic stopping mechanism with a `paper boat' potential\footnote{Another name for the Mexican-hat shaped potential characteristic of the Higgs scalar field.} of the relaxion is proposed in \citep{Wang_2019}. 
The model has many virtues compared to other relaxion models, but importantly, it is very ad hoc. 
The Higgs mechanism was charged with being ad hoc as it was only introduced to give the W and Z bosons mass. 
This relaxion model aims to resolve the problem by introducing a new very specific potential that would explain that other specific potential if it were just so. 
One ad hoc model is attempting to explain another. 
Each of the models in the relaxion-class of solutions brings one or more critical issues; resolving certain problems while introducing others and inevitably complicating the picture. 
The consistency issue with cosmological data is far from resolved, but rather than being a barrier to the pursuitworthiness, it is target and inspiration for future work.

The model's novelty and testability make it pursuitworthy. 
Models of axion-like particles ``are among the most compelling candidates for physics beyond the Standard Model (SM) of particle physics'' \citep[p.~1]{Choi:2020rgn}. 
The models represent an interesting and novel class of solutions to an existing problem. 
The models are also able to be tested and constrained by existing and future experiments. 
For example, there is sensitivity to the axion-nucleon couplings (via CASPEr and proton storage ring) and to the Higgs-axion mixing angle (from AION and AEDGE)\citep{Choi:2020rgn}. 
Of course, there is also cosmological data with which the models are rather poorly fit. 
The fact that there is conflict with experimental data that may be overcome means that there are targets and resources for continuing research and that progress can be gauged by this contact with experimental data.

\subsubsection*{Repulsive Gravity}		\label{subsub:grav}

Third, I will discuss the very aptly named repulsive gravity.\footnote{Repulsive gravity models are loosely related to the fifth force models and experiments discussed by \citet{Franklin2016-FRATRA}.} 
The hypothesis behind repulsive gravity is that there is a repulsive gravitational force that acts between matter and antimatter and that this could provide enough energy to drive cosmic expansion and explain the matter-antimatter asymmetry of the universe. 
As such, these models represents classes of possible solutions to the dark energy (DE) problem \citep{Hohmann_2010} and to the baryon asymmetry problem \citep{Noyes:1991yn}. 
Since the discovery of accelerated cosmic expansion in the late 1990s, cosmological models, such as $\Lambda$CDM, have had to include a term that represents negative pressure but is otherwise theoretically empty. 
The old idea of anti-gravity was given a new lease on life. 
Some of these models assumed that antimatter has negative gravitational charge \citep{Hajdukovic_2012}, in violation of the weak equivalence principle, or that it follows as a prediction of GR if the transformation of matter into antimatter requires not just a transformation of charge conjugation (C) but of parity (P) and time reversal (T) as well (i.e. CPT) \citet{Villata_2011,Villata_2012}. 

Some models are more baroque than others. 
\citep{Hohmann_2010} introduce $N \geq 3$ `dark' copies of the SM, each with its own gravitational metric. 
The metrics describe attractive gravitation within the groups, but repulsive gravity between them. 
The resulting matter-antimatter gravitational repulsion could explain the accelerated expansion of the universe while the additional matter would constitute the missing dark matter.
This model could potentially solve two long-standing problems, but suffers from a number of conceptual and empirical issues. 
It was nonetheless interesting enough for many physicists to pursue. 
It is, like the other models discussed above, very complex, and ontologically rich, proposing to triplicate\footnote{It triplicates, because the N=2 case was proven a no-go, so the model proceeded with $N \geq 3$.} all known particles. 
The model is externally inconsistent with a very widely accepted theory, viz. General Relativity. 
It had no empirical contact when it was proposed; sufficiently precise tests could not be conducted due to a lack of stable test antiparticles. 

This model is quite radical and general expectations for successful testing are were likely very low. 
Experimental tests require extreme precision and technical apparatus, but tests were recently conducted at CERN where antiprotons were already being produced as a matter of course. 
The ALPHA experiment at CERN has managed to make the production and capture of anti-hydrogen rather routine.
The ELENA ring can deliver $10^6$ antiprotons per minute, many of which can be cooled and trapped \citep{alpha_antimatter}.
The effect of repulsive gravity could be tested in a conceptually simple way: trap antiprotons in a container, open the top and bottom, and compare the number of annihilations out of each end, as demonstrated by an early proof-of concept experiment in 2013 following studies in \citep{Gabrielse:1989eer}.
Ordinary matter under the influence of gravity would predict that 80\% of the particles would escape through the bottom. 

Repulsive gravity was ruled out as evidence indicated that, indeed, antiprotons act just the same as protons under gravity. 
The experiment was able to conclude the following: 
``The probability that our data are consistent with the repulsive gravity simulation is so small as to be quantitatively meaningless (less than $10^{-15}$). Consequently, we can rule out the existence of repulsive gravity of magnitude 1$g$ between the Earth and antimatter. The results are thus far in conformity with the predictions of General Relativity. Our results do not support cosmological models relying on repulsive matter–antimatter gravitation.'' \citep{alpha_antimatter}. 
The negative result was strongly expected. 
The experiment was nonetheless conducted, because CERN already had anti-protons available and this hypothesis could be clearly tested with a straightforward experiment. 
It was pursuitworthy even though few had any hopes of this experiment would successfully overturn GR. 

\subsection*{Models of Lepton Flavour Universality Violation}

Lastly, I will briefly mention models of lepton flavour universality (LFU) violation. 
LFU is a feature of the standard model: the three generations of leptons are identical except for differences due to their masses.
The LHCb measured B meson decays where this universality seems to be violated with a $>2.6\sigma$ deviation from the SM \citep{Aaij_2015}, which increased to $3\sigma$ in 2021 \citep{Aaij_2021}.
This was below the threshold for statistical significance (5$\sigma$), but it was a tantalizing hint at a possible crack in the SM and spurred a great deal of activity (over 1300 citations). 
What was observed was that even though they ought to be produced at the same rate, there was an affinity for electrons compared to muons. 
It was theorized that this could be the result of new particles that have both lepton and quark properties (leptoquarks) or from new heavy particles like $Z'$ that would couple differently to electrons and muons. 
These models however well they could fit the anomalous data, were not well fit with other data. ``While the models may be easily adjustable to work with LFV in bottom hadron decays, they also have implications for other processes that cannot be accommodated easily and require a high level of fine tuning'' \citep[p.~138]{bechtleetal}.
Despite hopes for a direction in which to find new physics, more recent analyses have shown that the results are in line with SM expectations \citep{Aaij_2023}. 

There were various models that were proposed to solve this particular problem. 
They were most certainly easily testable and made direct contact with existing data. 
However, they lacked external consistency and in some cases even empirical validity beyond modelling this particular process, but these issues were bracketed in order to make progress on solving the LFU violation problem. 
These models were very narrow in scope, only purporting to solve a single issue and being not even applicable in a slightly broader domain. 

\subsection{Novelty, Testability, and Pursuit}	\label{sec:pursuit}

These models I presented are not cherry-picked as the ugliest models on the market, since those I probably would not have heard of. 
These are relatively popular models and are representative of the best and more pursuitworthy models particle physicists have to work on. 
As mentioned, I have called these models ugly because the traditional epistemic virtues that make theories beautiful, elegant, simple, intuitive, understandable, etc. are typically absent or meagre.
The models I presented: are ad hoc; have narrow empirical scope; are highly constrained; are fine-tuned; inconsistent with well-established models in other domains; not compatible with data in a broader scope; and compared to other models that have been ruled out are less simple, more ontologically expensive, more difficult to calculate, and not very plausible. 

What I mean in saying that these ugly models are pursuitworthy is that they should be worked on by some---it is reasonable to work on these models given our experimental context. 
Pursuitworthy models should instantiate the basic conditions of an adequate scientific model like being consistent enough to generate predictions and being mostly empirically adequate within a given domain (even though many pursuitworthy models have not yet had any empirical contact.)

Let us distil our lessons as clearly as possible. 
A pursuitworthy model must represent a class of potential solutions to an existing problem. 
In addition to this, there currently seem to be two features of the models we have examined that contribute most to their pursuitworthiness:
\begin{enumerate}[label=(\roman*)]
	\item the immediate and easy testability of the model or hypothesis
	\item the class of solutions represented by the model is radically novel
\end{enumerate}
These are not desiderata that should be maximized: the easiest testable hypotheses are likely trivial, and the most radically novel are probably incredibly unlikely. 
They are rather features that may constitute independently sufficient reasons for pursuit. 

By immediate and easy testability, I mean that a model i) makes clear predictions. It is sufficiently mathematically consistent and the values of parameters are sufficiently constrained such that it can be used to calculate the values of some observable quantities. ii) These predictions are in achievable energies and sensitivity or it might already be discoverable in existing or forthcoming data, or makes use of existing facilities (or with relatively small cost-effective amendments), very much along the lines discussed by \citet{Franklin1993-FRADPA-3}. 
Testability is then the possibility of constraining the empirical validity of the model. 
Some models may already be heavily constrained and so constraining them further may be less informative. 
The NMSSM can be constrained with existing LHC data; the relaxion star halos are observable and there are table-top axion experiments that have implications for the model; repulsive gravity was tested with antiprotons already a by-product of the LHC. 
Some hypotheses are being pursued for almost no other reason than that they can be easily tested, such as looking for magnetic monopoles in the discarded LHC beampipe simply because it was there \citep{Joergensen:2012gy}.
No one would have built the LHC in order to generate a used beampipe to test for an implausible hypothesis, but the cost was low and the potential payoff was high.
Contrary to what is widely said then, plausibility need play little role in pursuit if testability is high enough. 

I use the term `radical novelty' to distinguish from predictive novelty. 
The emphasis is less on the predictions the model makes than on its theoretical roots. 
By the class of solutions being radically novel, I mean that a model takes a genuinely new approach (rather than merely expanding on or continuing an old approach).
This virtue is extolled by the relaxion proposal, and to some degree by leptoquarks and repulsive gravity.
The NMSSM is a part of an existing class of supersymmetric models, but the relaxion made for such a pursuitworthy model because it in part introduced this new class of solutions. 
The relaxion makes new use of an existing concept---using an cosmological implications of the axion as a way to solve the Higgs naturalness problem is the kind of clever thinking that excited physicists and can kickstart a small research program.
Of course, what is novel is time-relative, and I am not trying to claim that the relaxion is more pursuitworthy than supersymmetry was. 
Rather, I mean to say that supersymmetry today is no longer a radically novel approach. 

Novelty represents a different approach to finding new physics than testability. 
Testability reflects the need to close the gaps, constrain parameter space, and make use of existing apparatus. 
Novelty reflects the need to take new approaches and reevaluate theoretical foundations, assumptions, and guides and come at problems from a different angle. 
Models that are highly novel may not be testable and may nonetheless be pursuitworthy and models that are highly testable but not so novel may also be pursuitworthy---this is what I mean when I say that they may constitute independently sufficient reasons for pursuit. 
Much theoretical work may need to be done before there can be any empirical access.
A great deal of creativity may be required to envision how to test the predictions of a model.
The achievable scope, scale, and precision of experimental physics lags far behind what theorists can dream of.  
Of course, unexpected empirical access may be opened up at any time, such as what happened with the studies of analogue black holes \citep{Barcel__2005}. 

Accounts, such as those of \citet{Franklin1993-FRADPA-3}, \citet{Laudan1977-LAUPAI}, \citet{McMullin1976-MCMTFO-3}, and \citet{Achinstein1993-PETHTD}, mostly picked up on at least six common criteria of pursuit. 
The first is often called the i) interest in the hypothesis, which indicates how interesting it would be to the community if the hypothesis were true. 
Second, there is ii) informativeness, which is how much one would learn (how much progress would be made) by pursuing the hypothesis. 
Third, there is the iii) plausibility of the hypothesis, which is that it is reasonable or is somewhat empirically supported. 
Fourth, there is the iv) ease of test, which has to do with how easily or clearly one can obtain a positive or negative result.
As Franklin points out, this includes both how easily one can distinguish the consequences of the hypothesis being true from it being false and how cost effective (including time) it would be to implement an actual experiment. 
Fifth, there is the v) analogousness of the hypothesis to past successes, as this may increase plausibility, or as it may facilitate model transfer (à la \citep{Nyrup2020-NYROWD}). 
Lastly, a model or hypothesis can be pursuitworthy if it has the conceptual resources for solving problems also known as its vi) heuristic power. 
All of these could indeed be reasons to pursue some model or hypothesis, but they are not all equally important. 
Judgments of pursuit will weight these (and other) criteria differently. 
These criteria may have vastly different weights in some contexts than in others and there is unlikely to be any general or principled reason that pursuit should more strongly determined by one rather than another. 
Let me go over these in turn in virtue of the case studies presented here. 

i) The interest-if-true would be enormous for any BSM model. 
A statistically significant discovery of a new effect (antigravity) or a new particle (relaxion or neutralino) would upend the current situation in particle physics, spurring a frenzy of model building, theorising and experimentation. 
``Any new hint would be a major discovery, whether it is the observation of a new particle, a new so-far unobserved phenomenon, or a non-trivial deviation from the standard model predictions''\citep[p.~292]{Abada2019FCCee}.
One only needs to look at the frenzy that followed the 2015 diphoton excess at 750~GeV to see how eagerly and enthusiastically physicists would welcome new physics.\footnote{For more about this response see \citep{RITSON20201}.} 
Because of this, it would be hard to say if any new physics discovery is more interesting than another. 
ii) The informativeness of the study estimates how much would be learned even if the experiment or study is not successful (otherwise it would get folded into (i) interest). 
This is realistically quite low for many of these models, since in the absence of success, a well-constrained parameter space only becomes more tightly constrained.
Consider the case of repulsive gravity where the negative result was highly expected. 
In doing the  theoretical work, one learns about the model, but not necessarily about the world as one confirms what one assumed to begin with. 
iii) Plausibility is rather hard to estimate as it is related to giving a probability of making a discovery. 
However, given the experimental evidence we have, it could be said that the plausibility of some of these models is quite low. 
Previously, searches for BSM physics would given much higher estimates of plausibility (prior to LHC null results), but the remaining ad hoc models in their tiny parameter spaces are not widely considered likely to succeed. 
Of course, the fact that there is a chance, even a small chance, at a major breakthrough is enough to be pursuitworthy. 
It is more of a possibility than a plausibility that counts. 
iv) With respect to ease of test, from what we have seen it almost cannot be stressed enough how strongly it constrains and guides what is pursuitworthy at the moment. 
If something is not easy or it is too expensive (in terms of various resources) then it will not likely be heavily pursued; if something is easy, one may pursue it regardless of how interesting, informative, or plausible it may be. 
v) The virtue of analogy is only likely to play a role where there is a tighter conceptual or mathematical analogy between an accepted and a novel application of model or idea, because the very idea that past successes will continue to work is something is being given up in pursuit of novel solutions. 
While, other things being equal this could be a benefit for the pursuit of some studies or models, it was not so relevant in the cases we examined, and rather, taking a new approach seemed to be more of a virtue. 
vi) Lastly, the problem-solving ability is certainly important insofar as models like the NMSSM come with tools that can be made ready to solve various problems and that classes of solutions inspire variants that may involve more clever thinking to avoid elimination. 
Being able to solve one problem is crucial, but solving more problems does not make something more pursuitworthy at the moment as these are harder to test.
The European Strategy for Particle Physics Update describes the situation as follows: ``On the theoretical front, less emphasis has been given to unified frameworks able to deal simultaneously with many key questions in particle physics, and more attention has been given to models that address individual shortcomings of the SM or simply single unexplained facts'' \citep[p.~137]{Ellis:2691414}.  
As a further remark, the \textit{rate} of problem solving that Laudan focuses on, is still not sufficient to capture what happens for these models we have looked at. 
They are in some cases merely proposing solutions to a single problem and creating further problems (see LFU models), and so the rate of problem solving is effectively 0. 

In the cases studied in this paper, some of these criteria play reduced roles and others play a stronger role than they did for other models in past decades. 
To round out the list of important criteria, we should add \textit{novelty}. 
Novelty has become important due to the current experimental context which would greatly benefit from novel approaches, if for no other reason than that past approaches have failed to turn up new physics.

One may wonder whether all I have said could be squared with a formal account, such as \citep{Nyrup2015-NYRHER}. 
The schematic nature of a formal model and the lack of ability to actual get any numbers out do not help understand changes like we have experienced in pursuitworthiness in particle physics. 
It is nice to envision in a simple calculus in which a term exists for informativeness (the expected epistemic value of pursuing a hypothesis given that that hypothesis is false), but it does not help us determine that value. 
I have here only provided a pluralistic sketch of pursuit and its experimental context dependence and I am reluctant to offer anything more formal.

One might be tempted to claim that these models are not really pursuitworthy and that they should not be pursued, perhaps arguing that this kind of HEP should be abandoned in favour of pure theory or devising ways of testing more `beautiful' models. 
When it comes to the normative part of the argument I want to argue that the response in adopting these two desiderata is a rational one in the given experimental context. 
This discussion necessarily brackets larger issues about whether one should pursue these models and experiments rather than focusing on an entirely different branch of physics. 
But this is not something an examination of pursuit should have to do. 
No account of pursuitworthiness I have come across has made a case for why a physicist should work on a particular model rather than take up biology or the flute. 
Nor do I intend to address whether the best course for discovering new physics lies in a new higher-energy accelerator or from letting deep neural networks loose on CERN data. 
My aim has been to demonstrate the rationality of pursuing ugly models given our experimental context in HEP. 

The normative dimension is operating at the scale of the discipline (not at the scale of governments, funding bodies, or individuals). 
There are no recommendations that all physicists should pursue the most testable and novel model. 
This is one area where pursuit and acceptance come apart: one can advocate the acceptance by all of the best empirically supported theory, but it is not rational to advocate that all should pursue the same model. 
Many avenues of research are rational to pursue.  
A reasonable division of labour at the level of the discipline is for some to work on beautiful high-energy theories and the foundations of QFT even if these cannot be readily tested, and others to pursue novel models that are fairly implausible, and others to perform precision measurements to squeeze parameter spaces even if little will likely be learned. 
I want to highlight the rationality of a multi-pronged approach to pursuing new physics and to highlight some of the experimental considerations for the reweighting of the `prongs'.

\section{Conclusion}

Novelty and testability are pursuitworthy qualities of models in our current situation and they can function as independent criteria for pursuit. 
Approaches based on maximizing theoretical virtues simply have not borne fruit and there are limited new experimental horizons available for the moment.
It makes a great deal of sense that these two virtues have come to the fore; these are two very reasonable responses to the situation where no new physics has been found and no new higher energy frontier is on the doorstep.
It may still be that these `ugly' models still exhibit some of the classic `beautiful' characteristics; all things being equal, a less complex model might be preferable if it was just as testable. 
But the fact remains that these ugly models do not exhibit theoretical virtues very much at all, and this alone means that we need to pay attention to pursuitworthiness in addition to acceptance. 
A logic of pursuit could only get us part of way, since, as I have shown, the context in which research taking place strongly determines in some cases what the dominant features of pursuitworthy models are. 
It has not always been the case, but in particle physics today, novelty and testability are extremely important and they should be. 
They can even compromise basic criteria like consistency and empirical adequacy.
In contexts where there is no forthcoming slew of new data and where old ideas have failed to open the door to new discoveries, pursuing models that are easy to test or that propose radical solutions to classes of existing problems is the reasonable course of action.

\bibliography{../../library}
\bibliographystyle{apalike}

\end{document}